\documentclass[a4paper,twocolumn]{revtex4-2}

\usepackage{graphicx}

\newcommand{\matP}{\mathcal{P}}

\begin{document}

\title{Simultaneous self-organization of arterial and venous networks driven by the physics of global power optimization}
\author{James P. Hague}.
\affiliation{School of Physical Sciences, The Open University, Walton Hall, MK6 3AY, UK}
\date{\today}

\begin{abstract}
Understanding of vascular organization is a long-standing problem in quantitative biology and biophysics and is essential for the growth of large cultured tissues. Approaches are needed that (1) make predictions of optimal arteriovenous networks in order to understand the natural vasculatures that originate from evolution (2) can design vasculature for 3D printing of cultured tissues, meats, organoids and organs. I present a method for determining the globally optimal structure of interlocking arterial and venous (arteriovenous) networks. The core physics is comprised of the minimization of total power associated with the whole vascular network, with penalties to stop arterial and venous segments from intersecting. Specifically, the power needed for Poiseuille flow through vessels and the metabolic power cost for blood maintenance are optimized. Simultaneous determination of both arterial and venous vasculatures is essential to avoid intersections between vessels that would bypass the capillary network. As proof-of-concept, I examine the optimal vascular structure for supplying square- and disk-like tissue shapes that would be suitable for bioprinting in multi-well plates. Features in the trees are driven by the bifurcation exponent and metabolic constant which affect whether arteries and veins follow the same or different routes through the tissue. They also affect  the level of tortuosity in the vessels. The method could be used to understand the distribution of blood vessels within organs, to form the core of simulations, and combined with 3D printing to generate vasculatures for arbitrary volumes of cultured tissue and cultured meat.
\end{abstract}

\maketitle

\section{Introduction}

In this paper, I provide a solution to the simultaneous prediction of arterial and venous structure by determining the optimal power (metabolic) cost of the whole vasculature, an important and longstanding problem of quantitative biology and biological physics. From a quantitative biology perspective, computer generation of vasculatures is needed to test hypotheses regarding the origins of vascular structures and their optimization during the evolution of organisms \cite{keelan2021}. In addition, the method described here fulfills unmet needs of regenerative medicine, engineered tissue, cultured meat, simulation, and biorobotics. Vasculatures provide the nutrient supply that is needed to increase the size of engineered tissue \cite{wang2021a} and cultured meat \cite{schatzlein2022}. For some applications, simulations may require vessels that are smaller than imaging resolutions, so computer generated vasculatures can be used to enhance those simulations \cite{hague2023a}. Bioprinted muscular tissues require dense vasculature if they are to be transplanted or used for biorobotics \cite{schatzlein2022}. Computational methods that find the vascular configuration by optimizing physical properties such as power can provide a solution to these needs.

Vascular networks are necessarily highly complex, since to provide nutrients to organs and tissues, they supply large numbers of capillaries on the microscale from small numbers of mm scale arteries \cite{graysanatomy}. This complexity is increased when the venous system is considered, as capillaries drain into venules which then combine into veins. The arteries supplying blood typically branch to become smaller arterioles that supply the tissue of organs. Blood exiting these organs then drains into venules that join to become major veins that return blood to the heart. Both the branching structures of arterioles and venules are connected to the mesh-like structure of capillaries \cite{graysanatomy}. Human capillaries have a diameter of approximately $5\mu\mathrm{m}$ \cite{POTTER198368}, whereas the largest arteries and veins can be around $10\mathrm{mm}$ in diameter \cite{graysanatomy}. This means that vessel size can vary by up to three orders of magnitude. Typically, networks of arteries and veins bifurcate to form a tree-like structure. The lumen size decreases exponentially as the tree is traversed due to this bifurcation.

Physics plays an important role in the organization of vascular trees: energy is expended both to pump viscous blood through the vasculature, and to maintain a volume of blood \cite{Murray1926a}. It is possible to determine a flow-radius relation for single vessels, $f\propto r^{\gamma}$ (known as Murray's law when $\gamma=3$), from the compromise between the power dissipated due to viscous flow and the metabolic cost for maintaining volumes of blood \cite{Murray1926a} (where $f$ is the flow, $r$ the vessel radius and $\gamma$ the bifurcation exponent). The optimization of power is much more complicated when large networks of vessels are investigated.

Quantitative questions remain about the role of optimization in the organization of large networks of vessels since there are vast numbers of ways to combine vessels to connect arteries and veins to the capillaries. As computational power has grown, it has become possible to investigate this combinatorial problem using a variety of methods. Approaches to predict the vascular structure in tissues and organs fall broadly into the following categories: (1) Algorithms that attempt to simulate angiogenesis (e.g. Refs. \cite{anderson1998a,mcdougall2006a}). (2) Algorithms that generate vasculatures based on local optimization principles \cite{Schreiner1993,Schreiner2006}. (3) Computational vascular growth taking into account the full global optimality of the tree \cite{keelan2016,keelan2019,keelan2021}. A scheme using a global tree pruning procedure is noted \cite{hahn2005}; but is not guaranteed to reach the global minimum. (4) Predictions based on morphological rules \cite{kaimovitzcoronary,Kassab1997}. (5) Multiscale methods (combining, for example, angiogenesis or local optimization with imaging information \cite{perfahl2011a,ii2020}). The algorithm described in this paper predicts vascular structure based on global optimality principles. The majority of algorithms for the generation of vasculatures concern single (i.e. arterial) vasculatures. 

The problem of organization in vascular networks has been investigated using a Simulated AnneaLing Vascular optimization (SALVO) approach for single vascular trees by Keelan {\it et al.} \cite{keelan2016, keelan2019, keelan2021}, who introduced a cost function for a single vasculature and determined the global minimum of that function. Simulated annealing is ideal for obtaining optimal computational solutions to problems with combinatorial complexity and is guaranteed to reach the global minimum for a sufficiently slow anneal. The cost function used in SALVO consists of a combination of the power cost associated with segments in the vasculature (similar to the considerations for deriving Murray's law \cite{Murray1926a}), in addition to penalties for vessels that penetrate tissue or hollow organs. The method has been used for computational growth of cerebral \cite{keelan2019} and cardiac \cite{keelan2016} vasculatures and to study deviations from Murray's law \cite{keelan2021}.

This paper builds on and goes beyond SALVO by predicting the structure of two intertwined vasculatures, such as arteries and veins, subject to penalties when arterial and venous vessel segments intersect. Any intersection between the two trees would divert flow around the capillaries, and, therefore, any network with an intersection would not be suitable for tissue supply. Algorithms to predict the structure of both arterial and venous networks suitable for tissue supply need to be capable of tracking all possible intersections between the trees. There are $N_1 N_2$ possible intersections, where $N_1$ and $N_2$ are the total number of segments in each tree respectively. Tracking the intersections adds significant computational cost and complexity to the problem of determining the optimal arrangement of vessels. I am not aware of any existing methods that track these intersections. For example, I note a multiscale dual vasculature approach based on a combination of imaging and local constrained constructive optimization\cite{ii2020}, which does not check for intersections between vasculatures or attempt to find the globally optimal vessel structure.

The paper is organized as follows: In section \ref{sec:methodology}, I discuss the development of a simulated annealing algorithm for multiple vasculatures. In section \ref{sec:results}, optimized vasculatures are presented for idealized tissue shapes suitable for growth in standard multi-well plates. In section \ref{sec:discussion}, I discuss prospects for the method. A summary and conclusions can be found in Sec. \ref{sec:conclusions}.

\section{Methodology}
\label{sec:methodology}

\subsection{Vascular tree representation}

In the approach presented here, the vascular tree is represented by straight segments representing vessels that join at nodes representing bifurcations. Two cases are considered: a single tree representing arteries only and a two-tree arrangement representing arteries and veins. Within the arterial tree vessels bifurcate in the direction of flow. In the venous tree vessels join in the direction of flow. No anastomoses are included.

The following terminology will be used: In the arterial tree, a node is supplied by a parent segment and drains into child segments (the opposite being the case in the venous tree), such that the parent segment is larger than the child segments. The node at the end of the parent segment is referred to as the parent node and the nodes at the end of the child segments as child nodes. The two child segments of a node are sibling segments to each other. The smallest nodes in the tree are referred to as leaf nodes. For a schematic representation see e.g. Ref. \onlinecite{keelan2021}.

Leaf nodes (the smallest nodes) of the tree are positioned using a Poisson sphere algorithm. This ensures that nodes have approximately even spacing, regardless of the shape of the tissue. Both trees share the same leaf-node positions to ensure flow from the smallest arterioles of the first tree into the smallest venules of the second. Leaf-node positions are not changed during the optimization, reducing the overall number of node positions that need to be optimized by the algorithm. 

The algorithm can handle any tissue shape and two tissue shapes are considered here: (1) A disc with radius $R=8\mathrm{cm}$ and thickness $\Delta=2\mathrm{cm}$, with vessels entering from the center of the disk at the bottom (coordinates $-1\mathrm{cm},0,-2\mathrm{mm}$ and $-1\mathrm{cm},0,2\mathrm{mm}$). (2) A square tissue with side $A=\mathrm{8cm}$ and thickness $\Delta=2\mathrm{cm}$ with vessels entering from the corner of the square (coordinates $-4\mathrm{cm},-4\mathrm{cm},-2\mathrm{mm}$ and $-4\mathrm{cm},-4\mathrm{cm},2\mathrm{mm}$). The tissue shape is used as part of the Poisson disk process during initialization and provides constraints for the placing of other nodes during the optimization (see Sec. \ref{sec:methodology-tissue-contraint}).

\begin{figure}
    \centering
    \includegraphics[width=60mm]{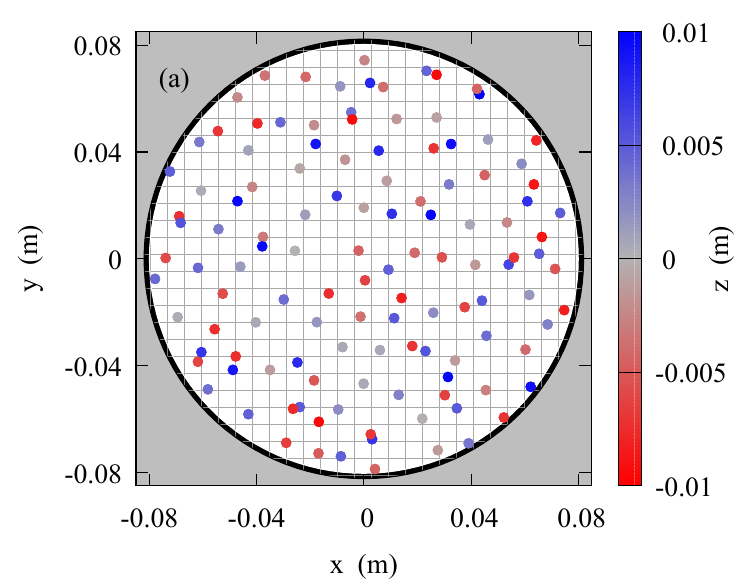}
    \includegraphics[width=60mm]{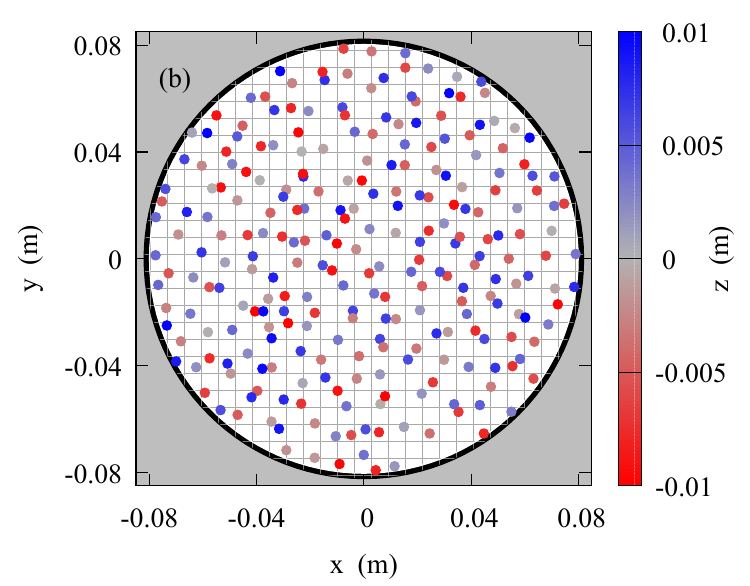}
    \includegraphics[width=60mm]{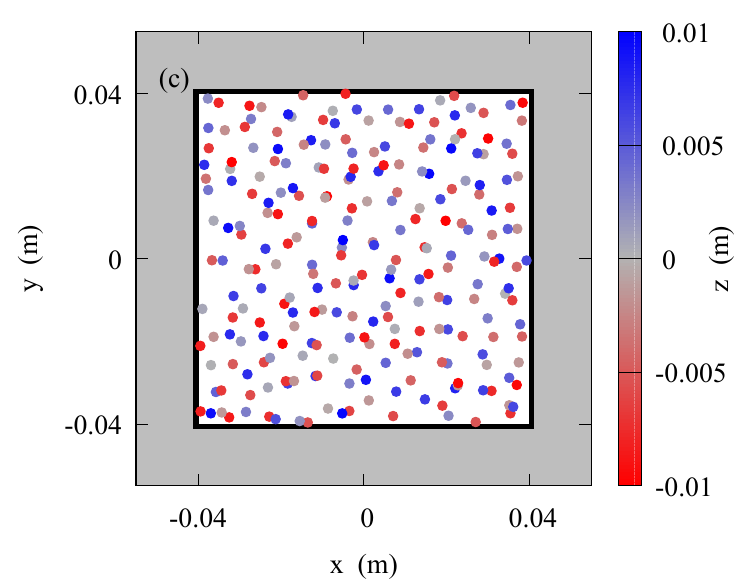}
    \caption{Distribution of leaf nodes by the Poisson sphere algorithm for (a) a disc shaped well of tissue and 122 leaf nodes (b) a disc shaped well with 255 leaf nodes (c) a square well with 256 leaf nodes. The Poisson sphere algorithm fills the shape with well-distributed points.}
    \label{fig:tissuepoissonsphere}
\end{figure}

Examples of leaf-node positions for the disc-well-plate and square well shapes can be found in Fig. \ref{fig:tissuepoissonsphere}. Square tissue shapes have been used to study deviations from Murray's law in single vascular trees in two dimensions \cite{keelan2021}. Due to the differences in shape between the tissues, the number of leaf nodes is slightly different for the circular and square tissue shapes. This does not affect the algorithm and the number of leaf nodes is shown on each plot.

In the following, the terms ``two tree'' and ``multiple vasculatures'' will be used to denote the calculations including arterial and venous trees. Whereas ``single vasculature'' and ``one tree'' will denote a calculations involving a single tree representing the arterial network only.

\subsection{Cost (objective) function}

The goal is to minimize an objective function for the vasculature based on the total power cost for blood flow through the vasculature and maintaining the vasculature, subject to penalties that act as constraints for the vasculature. Details of the objective function can be found in this section.

\subsubsection{Power cost (metabolic cost)}

There are two contributions to the metabolic cost of a vasculature. The first relates to the metabolic cost to maintain a volume of blood and the second to the power cost to pump blood through the tree. These are summed to determine the total metabolic cost:
\begin{equation}
  P = \sum_{j} \left( m_b \pi r_{j}^2 l_{j} + \frac{8 \mu f_{j}^{2} l_{j}}{\pi r_{j}^{4}}\right)
  \label{eq:singlesegment}
\end{equation}
where the subscript, $j$, is an index representing the segment, $r_{j}$ is the segment radius, $l_{j}$
its length, $f_{j}$ its volumetric flow. The metabolic power
cost of blood is $m_b$. The dynamic viscosity of blood is $\mu$. If the radius is small, costs associated with viscosity dominate, but the costs associated with maintaining blood become tiny. On the other hand, a very large radius leads to tiny costs associated with the flow through the segment, but the cost of maintaining blood becomes huge. This leads to a compromise where the optimal radius is of intermediate size.

Equation \ref{eq:singlesegment} can be rearranged into the form,
\begin{equation}
  P = m_b \pi r_{\rm root}^2 \left( \sum_{j} \left(\frac{r_{j}}{r_{\rm root}}\right)^2 l_{j} + \frac{8 \mu f_{\rm root}^2 (f_{j}/f_{\rm root})^{2} l_{j}}{\pi^2 m_{b} r_{\rm root}^6 (r_{j}/r_{\rm root})^{4}}\right)
  \label{eq:singlesegmentrescaled1}
\end{equation}
and then for compactness, the metabolic ratio $\Omega_{\rm root} = m_b\pi^2 r_{\rm root}^6/8\mu f_{\rm root}^2$ may be introduced \cite{keelan2021} to obtain,
\begin{equation}
  P = m_{b}\pi r_{\rm root}^2 \left( \sum_{j} \left(\frac{r_{j}}{r_{\rm root}}\right)^2 l_{j} + \frac{ (f_{j}/f_{\rm root})^{2} l_{j}}{\Omega_{\rm root} (r_{j}/r_{\rm root})^{4}}\right)
  \label{eq:singlesegmentrescaled2}
\end{equation}

Note that the metabolic ratio for the root node is used here, since the root flow and radius are fixed.

\subsubsection{Self-avoidance penalty}

When arterial and venous vascular networks are simultaneously optimized, it is essential that the resulting vasculatures do not intersect, except with their directly neighboring segments (i.e. other segments in a bifurcation or the common segments attached to leaf nodes). Thus, intersections between the two vascular trees need to be penalized in addition to penalizing intersection of a vasculature with itself. Segments that join at a node are not included in the penalty, since they must intersect by definition. For each segment this includes the two child segments, a sibling segment and a parent segment; with the exception that segments attached to leaf nodes intersect with their parent and sibling segments within the same tree, and the segment in the other tree attached to the same leaf node.  It is noted that this penalty has not been used in previous single vasculature optimizations using SALVO \cite{keelan2016,keelan2019,keelan2021}: in those cases such 
a penalty would remove anastomoses in principle, although in practice these were not found to be a problem with single trees.


Vessels are considered to be too close (intersected) if,
\begin{equation}
    D_{ij} < D_{ij,\rm avoid} = \alpha (r_{i}+r_{j}).
    \label{eqn:intersectioncondition}
\end{equation}
where $D_{ij}$ is the closest approach distance between segments. Here the constant, $\alpha$, is chosen to be $\alpha=2$. $r_{i}$ and $r_{j}$ are the radii of the segments. 

If Condition \ref{eqn:intersectioncondition} is satisfied, the following penalty is applied:
\begin{equation}
C_{\rm avoid} = P_{\rm avoid} \sqrt{1-D_{ij}/D_{ij,\rm avoid}}
\end{equation}
In this way, the penalty is introduced smoothly to assist the optimization algorithm to push apart intersecting segments. The constant $P_{\rm avoid} = 0.1\mathrm{J s}^{-1}$ is chosen. For well annealed vasculatures, this avoidance penalty is sufficiently large that no intersections are recorded in the final vasculature.

\subsubsection{Tissue constraint}
\label{sec:methodology-tissue-contraint}

A penalty to keep nodes within the tissue shape is also applied. It has the form,
\begin{equation}
    C_{\rm tissue} = P_{\rm tissue} N_{\rm out}
\end{equation}
where $N_{\rm out}$ is the number of nodes outside the tissue shape and $P_{\rm tissue}$ is a large constant with units of power.

In the calculations presented here, nodes are constrained to move within tissue shapes formed of simple contiguous solids.  During the optimization, any moves that place nodes outside the tissue are rejected, consistent with $P_{\rm tissue}\rightarrow\infty$. 

\subsubsection{Total cost (objective) function}

All these factors are combined to make a total cost (or objective) function of:
\begin{equation}
    C_{\rm total} = P + C_{\rm avoid} + C_{\rm tissue}
\end{equation}
The aim is to find the configuration of vessels with the minimum value of the total cost function.

\subsection{Simulated annealing}

Simulated annealing is a powerful technique for finding the global minimum of a cost function. It has several advantages over other optimization techniques, such as ease of implementation and guarantee of finding the global minimum. 

On each iteration of the simulated annealing algorithm, updates are made that change the configuration of the vascular tree. The configuration being both the positions of the nodes and the way that nodes are joined. The number of ways that nodes can be joined together grows factorially (combinatorially) with tree size. Simulated annealing works well for combinatorial problems (e.g. the traveling salesman problem), making it a good choice for the combinatorial problem of vascular networks described here.

Five distinct types of updates are made during the anneal. The minimal set of updates that ensure ergodicity involve (1) moving the position of a single node and (2) swapping the parent nodes of two nodes selected at random \cite{keelan2016}. There are several additional updates used here that lead to rapid traversal of the tree configuration space. These are chosen with various probabilities summarized in Table \ref{tab:updates}, and described as follows: (3) In addition to moving a single node, it is possible to move a whole segment such that it does not change orientation (i.e. to move start and end nodes of the segment in parallel). (4) Furthermore, four nodes selected at random may be moved simultaneously. (5) Finally, it is also possible to select a node and then to reconnect the sub-tree of that node (i.e. all downstream nodes in the arterial tree, or all upstream nodes in the venous tree). By reconnect, it is meant that the nodes are not moved, but that a new topology (new set of connections between nodes) is selected for those nodes at random. By making large changes to tree topology, reconnection updates have low acceptance rate but high impact.

\begin{table}[]
    \centering
    \begin{tabular}{|p{0.8\linewidth}|c|}
    \hline
    Update & Weight\\
    \hline
Move a single, randomly selected, node & 53.57\%\\
Move a randomly selected segment & 35.71\%\\
Move four randomly selected nodes & 4.46\%\\
Swap parents of two randomly selected nodes & 4.46\%\\
Reconnect sub tree of randomly selected node &  1.79\%\\
\hline
    \end{tabular}
    \caption{Updates types and the frequency with which they are selected.}
    \label{tab:updates}
\end{table}

\begin{figure}
    \centering
    \includegraphics[width=85mm]{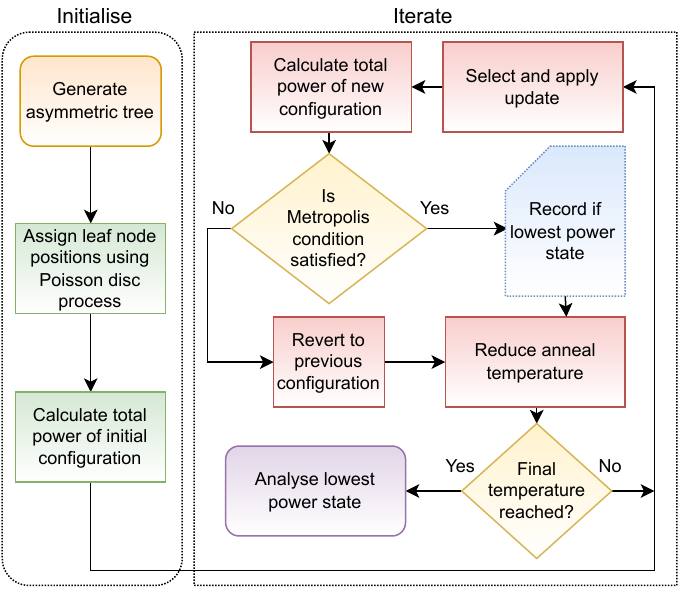}
    \caption{Flow chart of the algorithm, reproduced from Ref. \cite{keelan2021}. The main algorithmic difference here is that updates involve two trees and an extended set of updates is used. The objective function also differs through the inclusion of penalties to suppress any intersections between venous and arterial trees.}
    \label{fig:flowchart}
\end{figure}

At the heart of simulated annealing is the Metropolis condition, which favors downhill optimization, while also allowing for uphill searching to avoid falling into local minima. Once the configuration of the tree is changed according to the chosen update, the new configuration is either accepted or rejected according to the probability,
\begin{equation}
\matP_{\theta,\theta+1} = {\rm min}\left\{\exp\left(\frac{-\Delta C_{\rm total}^{(\theta,\theta+1)}}{T_{\theta}}\right),1\right\}\label{annealing}
\end{equation}
where $\theta$ represents the iteration number, $\Delta C_{\rm total}$ the difference in cost function due to the configuration change and $T_{\theta}$ the anneal temperature on iteration $\theta$. An exponential temperature schedule is used here, where $T_{\theta+1}=(1-\delta)T_{\theta}$, such that,
\begin{equation}
    T_{\theta} = (1-\delta)^{\theta} T_{0}
\end{equation}
and $\delta\ll 1$ is a small number that represents the anneal rate. $T_{0}$ is the initial temperature. As $\delta$ is decreased, the length of the anneal increases. Note that $T$ is not the actual temperature of the tissue, rather the anneal temperature dictates how much uphill searching is carried out: for low $T$ towards the end of the anneal the search is mainly downhill and towards the start of the anneal high $T$ leads to similar uphill and downhill searching. A flowchart outlining the algorithm is shown in Fig. \ref{fig:flowchart}.

\section{Results}
\label{sec:results}

\begin{table}[]
    \centering
    \begin{tabular}{||c||c||c|c|c|c||}
    \hline
        name & symbol & \multicolumn{4}{|c||}{value} \\
        \hline
        \hline
        anneal parameter & $\delta$ & \multicolumn{2}{|c|}{$3\times 10^{-4}$} & \multicolumn{2}{|c||}{$1\times 10^{-4}$} \\
        \hline
        metabolic constant & $m_{b}$ & \multicolumn{4}{|c||}{$641.3\,\mathrm{J\, s}^{-1}\mathrm{m}^{-3}$} \\
        \hline
        blood viscosity & $\mu$ & \multicolumn{4}{|c||}{$3.7\times 10^{-3} \,\mathrm{Pa \,s}$}\\
        \hline
        root radius & $r_{\rm root}$ & \multicolumn{4}{|c||}{$1\times 10^{-3}\,\mathrm{m}$} \\
\hline
        root flow $\, (\mathrm{ml\, min}^{-1})$ & $f_{\rm root}$ & $39.24$ &  $27.75$ &  $19.62$ & $5.0$\\
        \hline 
        root metabolic ratio & $\Omega_{\rm root}$ & $0.5$ &  $1$ & $2$ & $30.8$ \\
        \hline
    \end{tabular}
    \caption{Input parameters. Note that $f_{\rm root}=\sqrt{m_{b}\pi^2 r_{\rm root}^6/8\mu \Omega}$.}
    \label{tab:my_label}
\end{table}

\begin{figure*}
    \centering
\includegraphics[width=80mm]{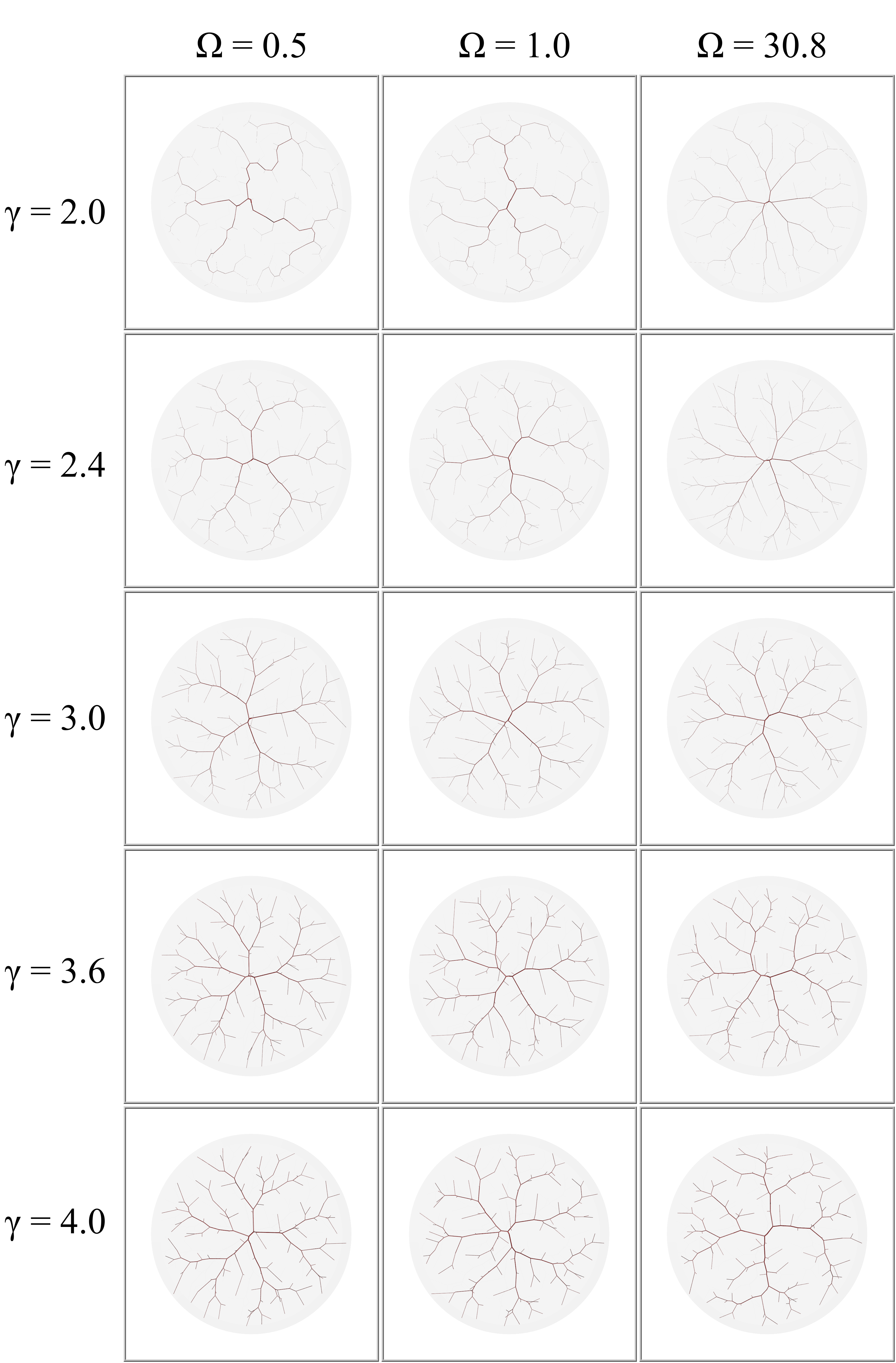}
\includegraphics[width=80mm]{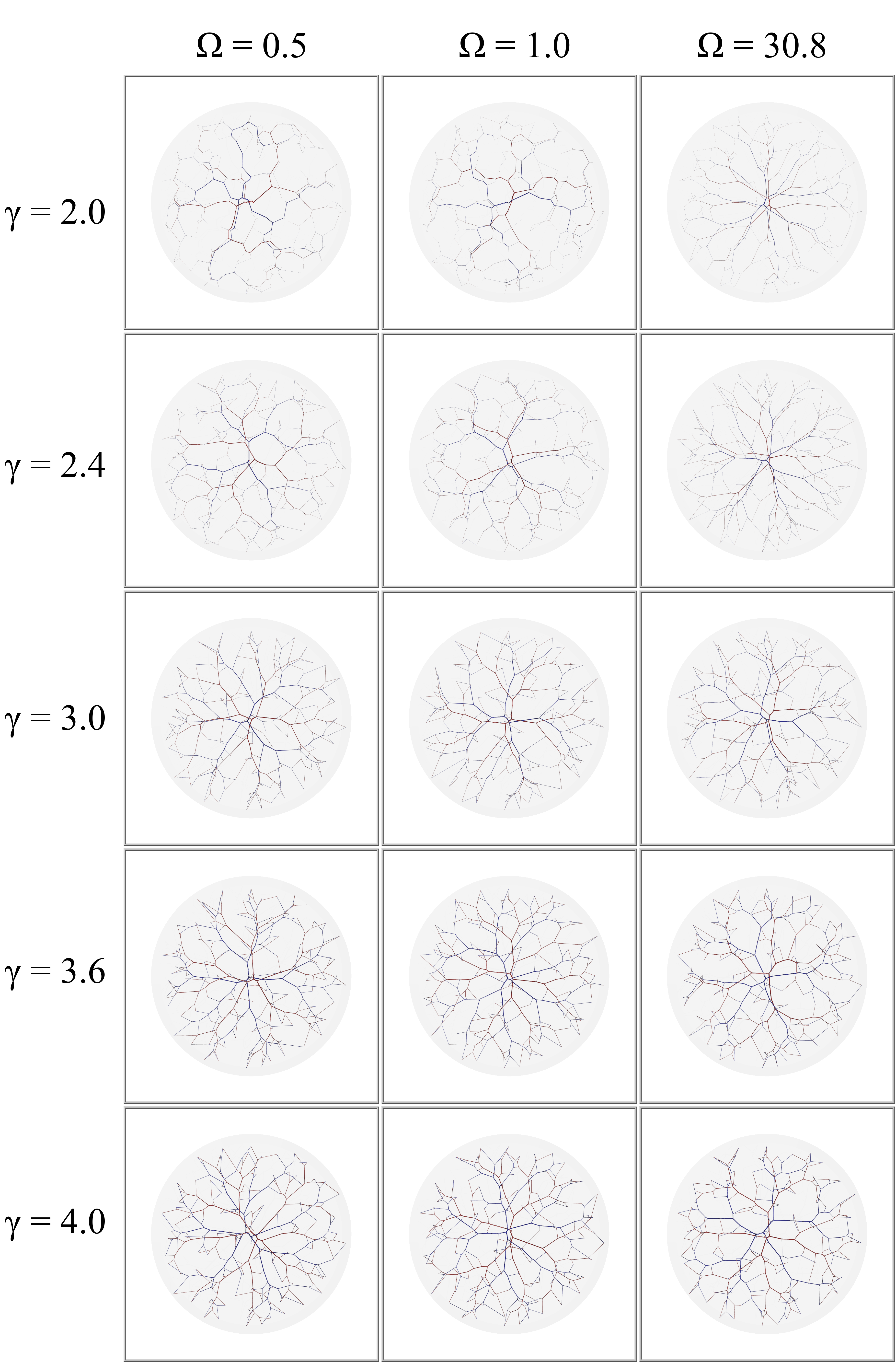}
    \caption{Optimized vessels for a thin disc of tissue. $N=122$. The optimized single tree is shown in red. Results for two simultaneously optimized trees are shown with the individual trees identified in red and blue. For single trees with small $\Omega$ and $\gamma$, vessels are tortuous, and for large $\Omega$ and small $\gamma$ vessels are elongated. When two sets of vessels are included, the structure of the individual trees is similar to the one tree case, but the correlation between the paths of the vessels depends on the specific values of $\Omega$ and $\gamma$. For the very tortuous vessels identified for $\gamma=2$ and $\Omega=0.5$ there is little correlation between the paths of the arterial and venous trees, and for large $\gamma$ the paths of venous and arterial trees are uncorrelated. Straight vessels from the two trees tend to follow each other. For example, when $\Omega=30.8$ and $\gamma=2$ the vessels are very straight and both trees follow very similar paths (while still avoiding each other). Trees are moderately correlated in other cases.}
    \label{fig:disc128}
\end{figure*}

Figure \ref{fig:disc128} shows the pattern of optimized vessels for a well plate (disc) arrangement supplying $N=122$ leaf nodes. Patterns for a single arterial tree (red only) and combined venous and arterial trees (red and blue vessels) are shown. For single trees with small $\Omega$ and $\gamma$, vessels are tortuous. For large $\Omega$ and small $\gamma$ vessels are elongated. Input vessels are located in the center of the lower side of the disc. 

When two sets of vessels are included, the correlation between the paths of the two trees depends sensitively on values of $\Omega$ and $\gamma$. For the very tortuous vessels identified for $\gamma=2$ and $\Omega=0.5$ there is almost no correlation between the paths of the vessels, as is also the case for the largest $\gamma$ considered here. The straighter the vessels, the more correlation can be seen between the vessels in the two trees. For example, for $\Omega=30.8$ and $\gamma=2$, the vessels are very straight and the paths are correlated. For other cases the vessel locations in the trees are moderately correlated. Vessels from the two trees do not intersect, so that anastomoses and short circuits between the two trees are not present (note that vessels are able to pass over or under each other due to the thickness of the discs). The structure of the two trees is similar to the one tree case. In physiological systems (e.g. the liver) where the vessels enter at the same location it is common for vessels to follow similar paths through the tissue \cite{ABDELMISIH2010643}.

\begin{figure*}
    \centering
\includegraphics[width=82mm]{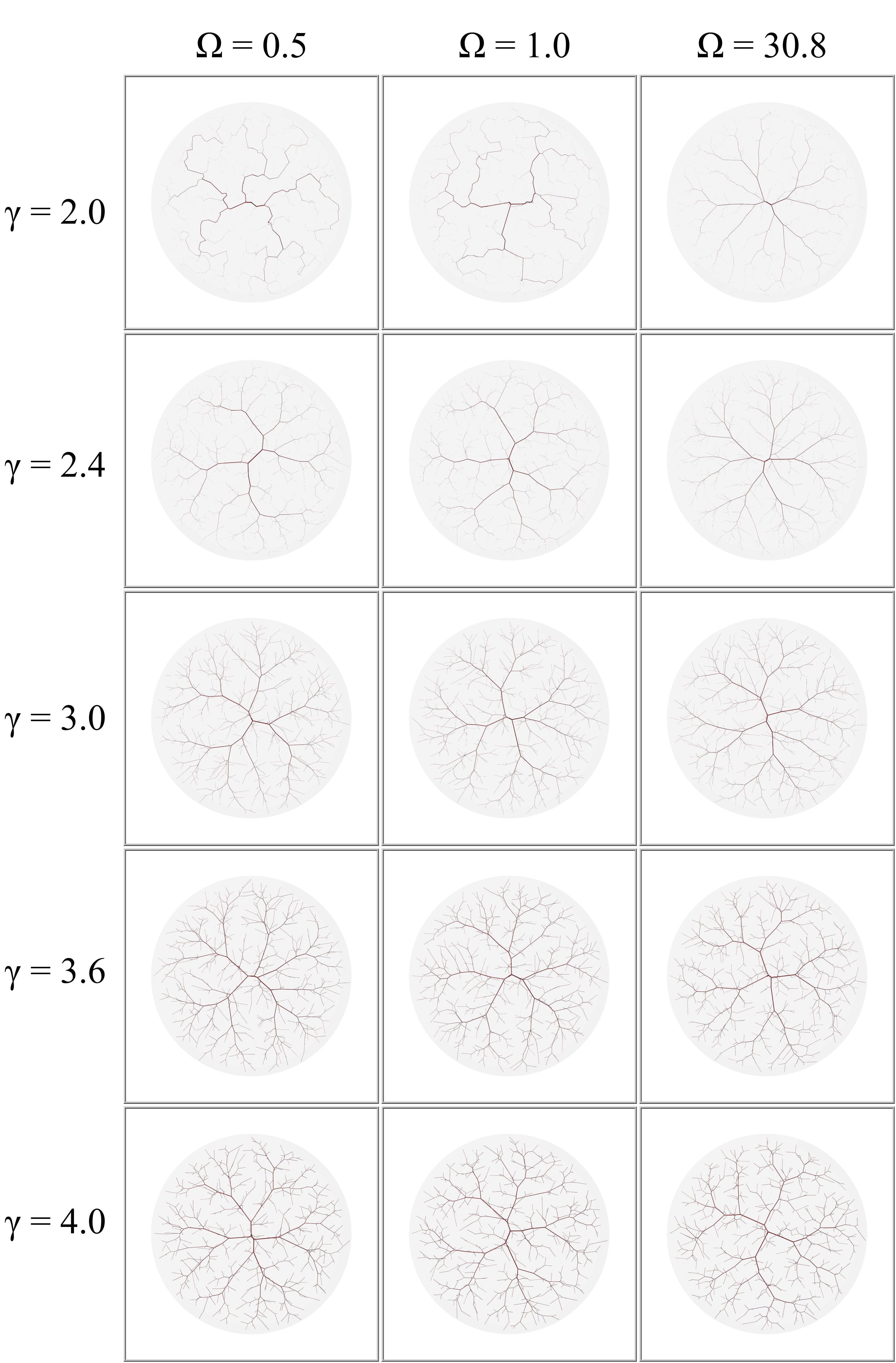}
\includegraphics[width=82mm]{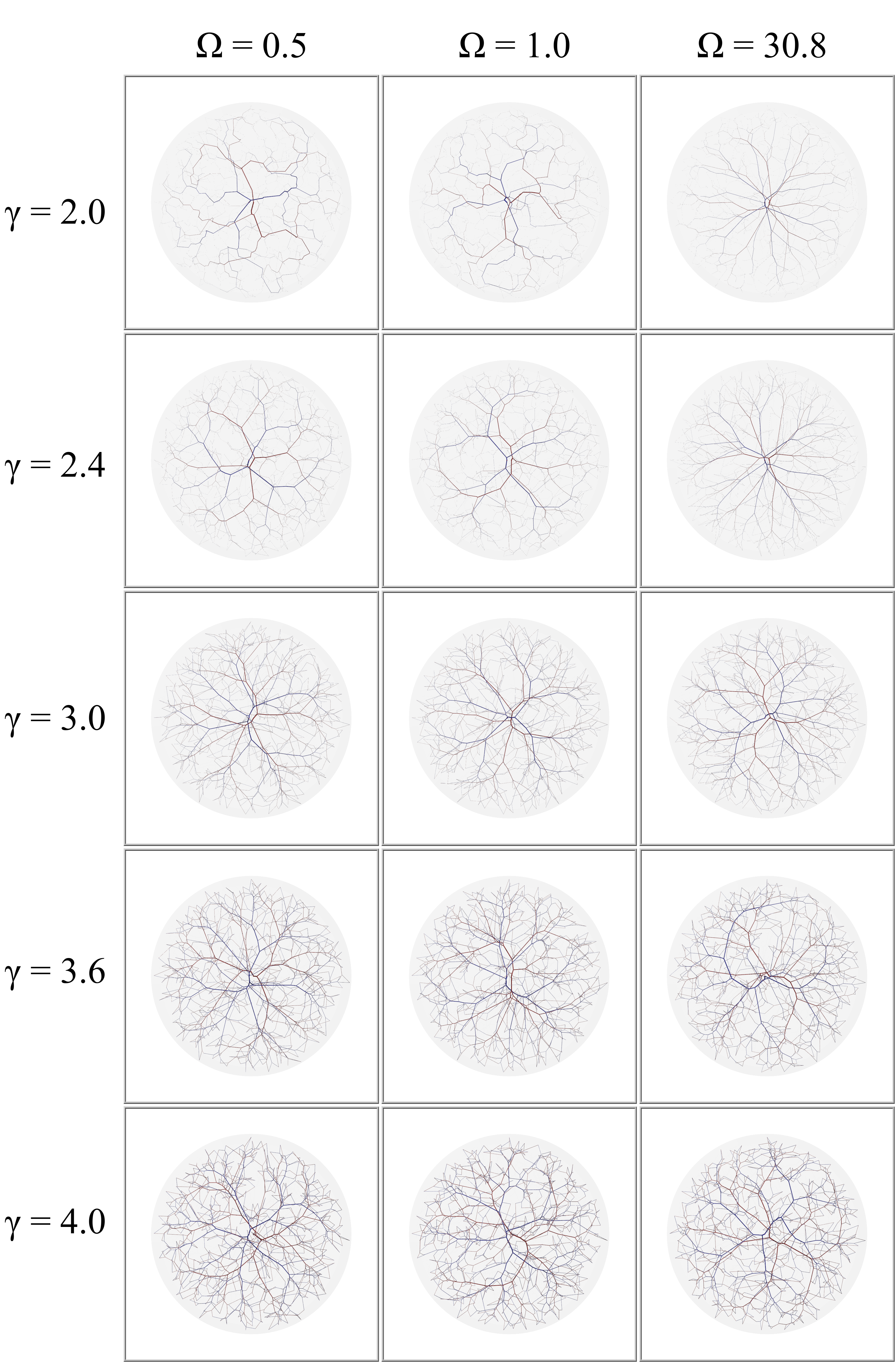}
    \caption{The arrangement of optimized vessels for a thin disc of tissue with $N=488$ shows that the tree structure is not sensitive to the number of leaf nodes, which mainly dictate the level of detail in the vascular structure. Optimization of a single tree is shown in red. Optimization of two trees shown in red and blue.}
    \label{fig:disc512}
\end{figure*}

The tree structure is not sensitive to the number of leaf nodes, which mainly dictate the level of detail in the vascular structure. Similar results to the $N=122$ leaf-node case are found for well-plate shaped tissues with $N=255$ (not shown) and $N=488$ leaf nodes (shown in Fig. \ref{fig:disc512}). With an increase in leaf nodes, the overall structure of the largest vessels remains similar, while more detail is seen in the smallest arterioles and venules.

\begin{figure*}
    \centering
 \includegraphics[width=75mm]{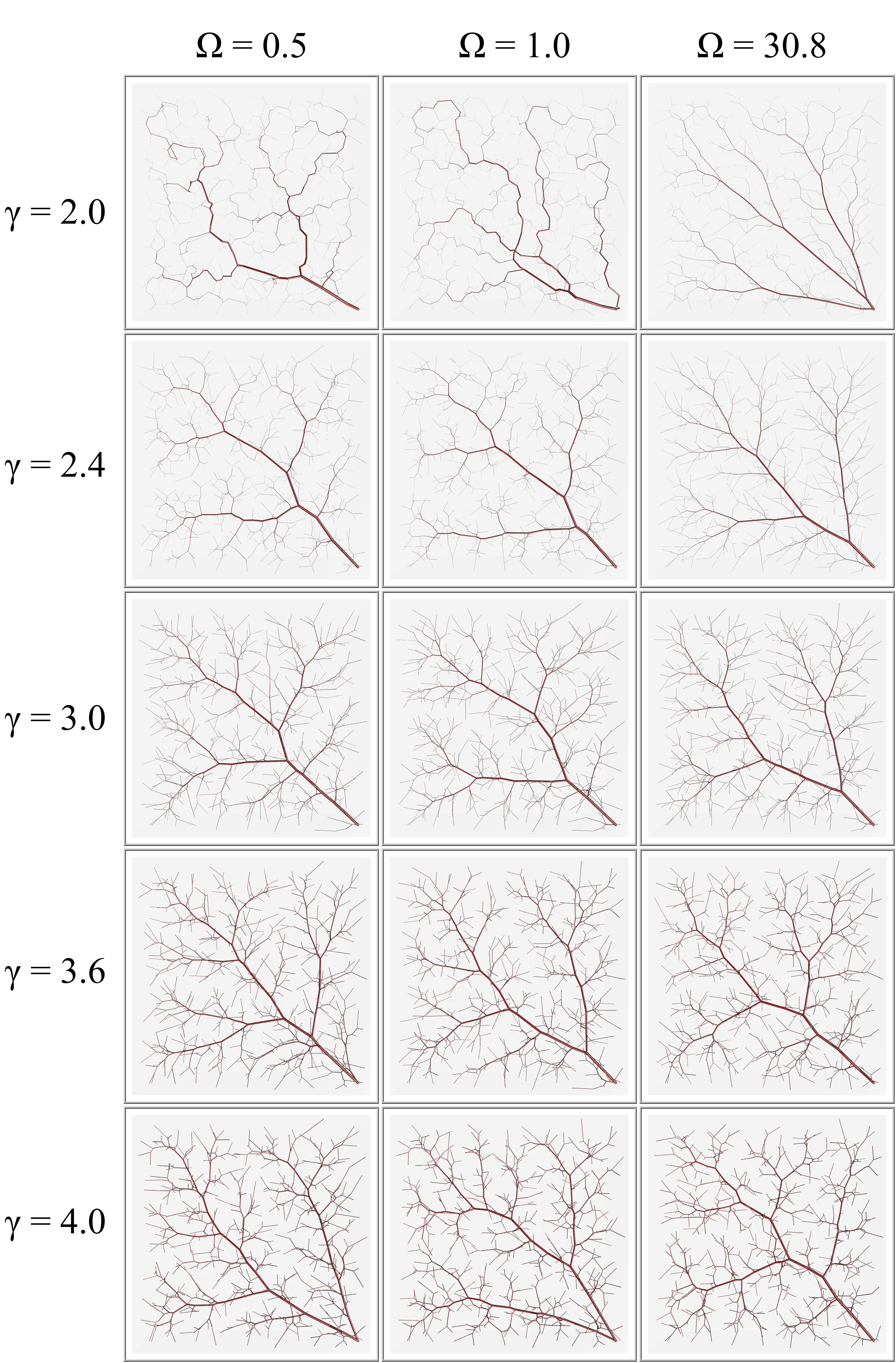}    
  \includegraphics[width=75mm]{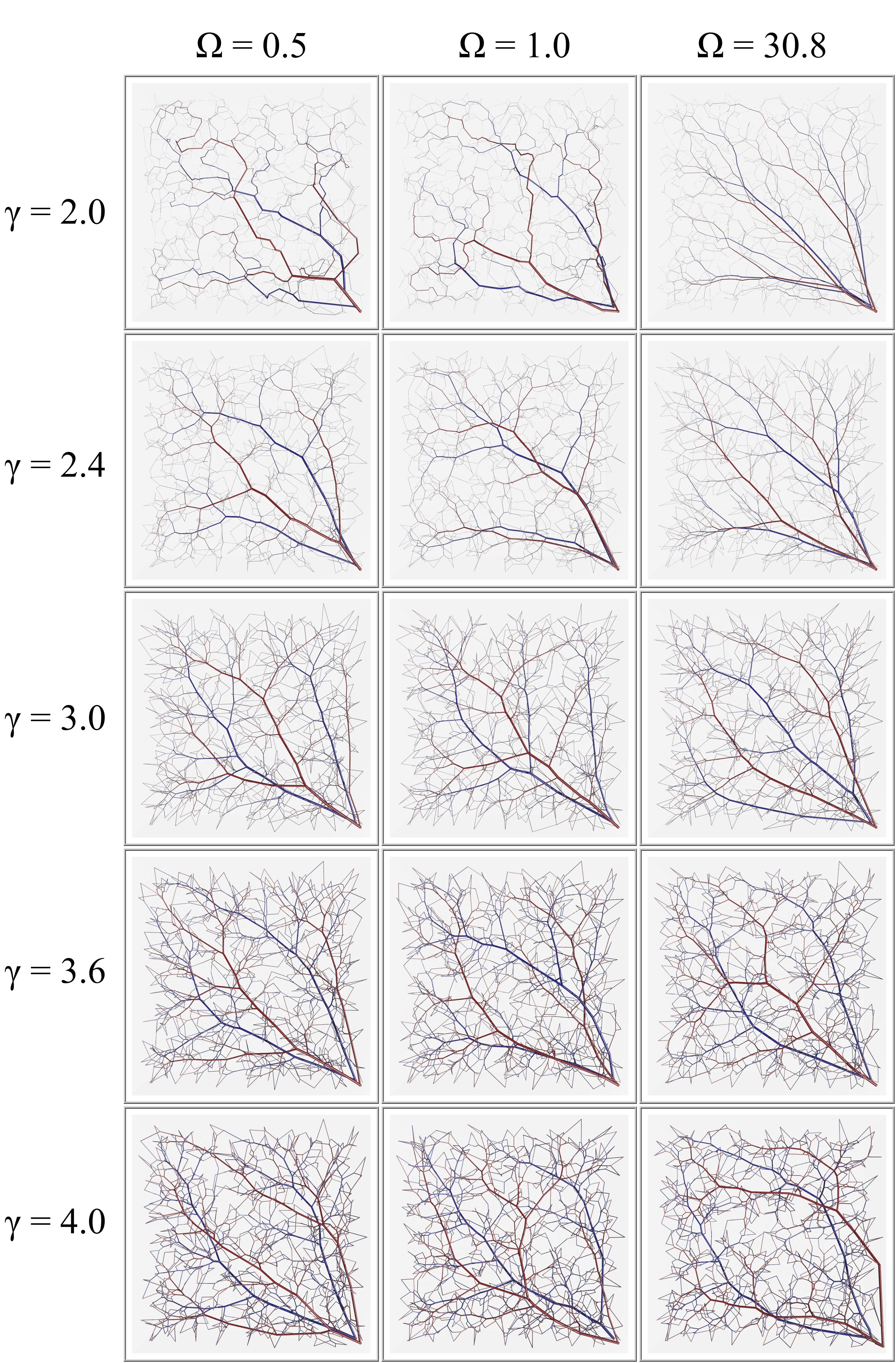}    
    \caption{Vessels optimized for thin square tissues with $N=506$ show that the detailed arrangement of vessels varies with tissue shape, whereas the tortuosities of vessel paths for the $\Omega$ and $\gamma$ considered are similar in both disc and square shaped tissues. Vessels for single trees are shown in red. Optimized vessels for two trees are shown in red and blue. }
    \label{fig:square512}
\end{figure*}

The detailed arrangement of vessels varies with tissue shape, whereas properties such as tortuosities of vessel paths for the different $\Omega$ and $\gamma$ are similar in different tissue shapes. Fig. \ref{fig:square512} shows optimized vessels for a thin tissue arrangement with a square shape. Vessels enter from the lower right corner. As in the case of the thin disc of tissue, the vessels are tortuous for $\gamma=2$ and small $\Omega$, whereas they are long and straight for $\gamma=2$ and $\Omega=30.8$. For $\gamma=3$, all trees look similar and the structure is largely independent of $\Omega$. As $\gamma$ increases, the trees become more branched. The first large branch is situated closer to the root node for small $\Omega$ and further from the root node for large $\Omega$. As the number of trees is increased to two within the thin square of tissue, the vessels in both trees follow similar paths when $\Omega$ is largest. Whereas, for the smallest $\Omega$ and small $\gamma$ the tortuous paths obtained tend to follow different routes.

\begin{figure*}
    \includegraphics[width=75mm]{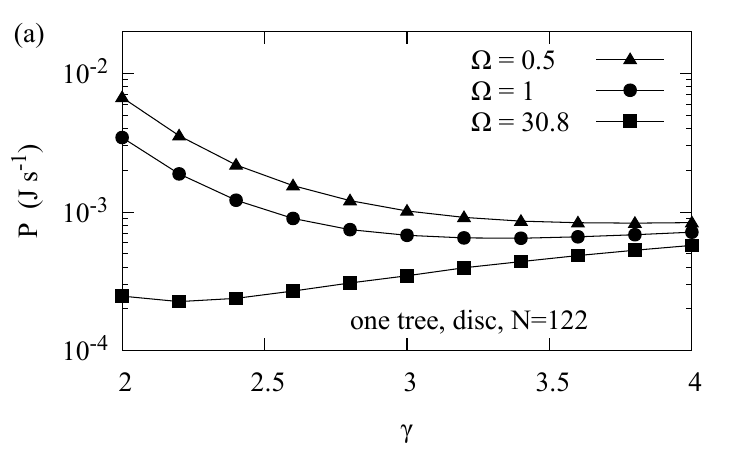}
    \includegraphics[width=75mm]{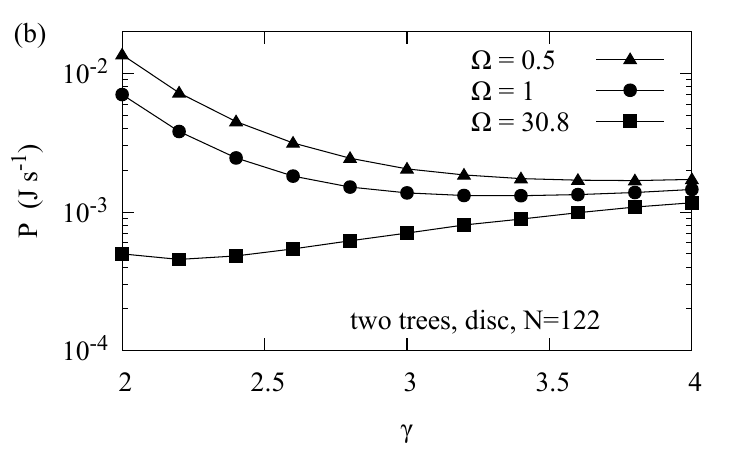}
    \includegraphics[width=75mm]{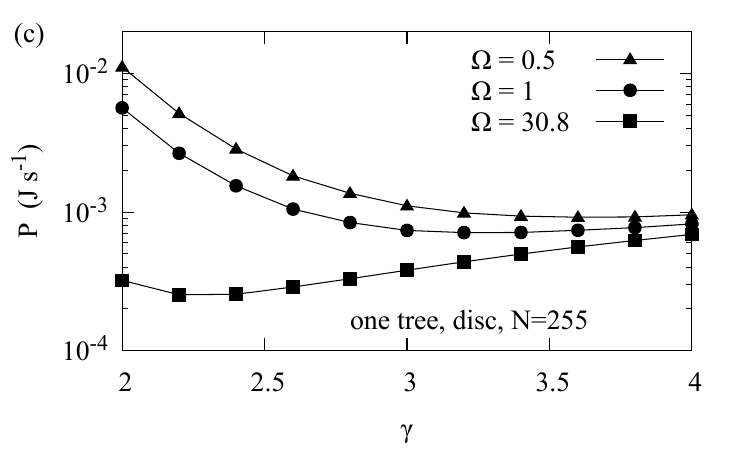}
    \includegraphics[width=75mm]{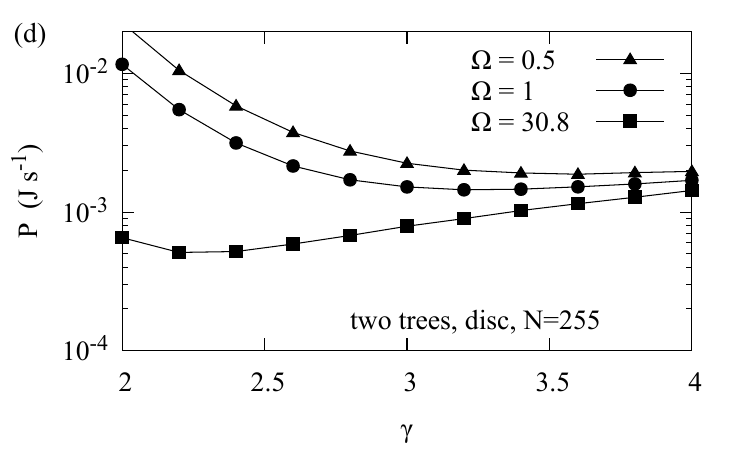}
    \includegraphics[width=75mm]{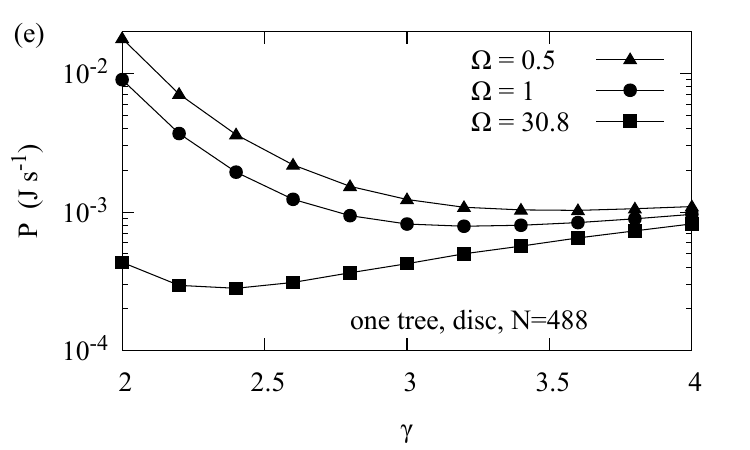}
    \includegraphics[width=75mm]{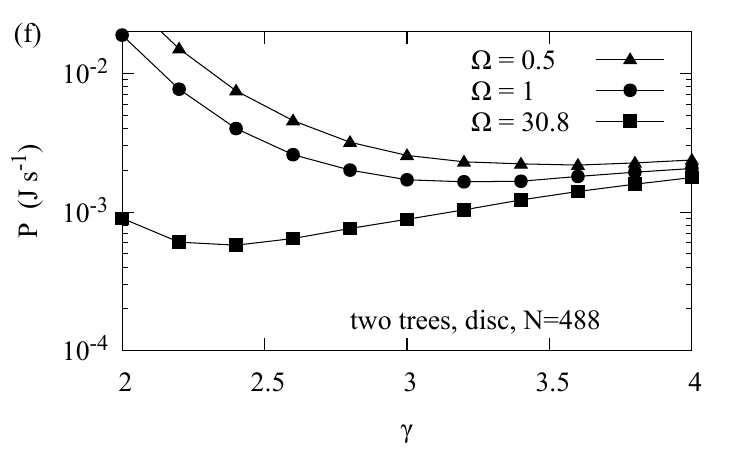}
    \caption{Metabolic power demand vs $\gamma$ for a disc of tissue to identify optimal $\gamma$. The minimum in the metabolic power, $P$, vs $\gamma$ depends strongly on $\Omega$, but only weakly on $N$, and is essentially independent of the number of trees. The structures of the relationships are similar for different tree sizes and both single and dual tree vasculatures. Panels show results for (a) a single tree, 122 leaf nodes (b) arterial and venous trees, 122 leaf nodes (c) a single tree, 255 leaf nodes (d) arterial and venous trees, 255 leaf nodes (e) a single tree, 488 leaf nodes (f) arterial and venous trees, 488 leaf nodes. }
    \label{fig:powerdisc}
\end{figure*}

The minimum in the metabolic power, $P$, vs $\gamma$ depends strongly on $\Omega$, but only weakly on $N$, and is essentially independent of the number of trees. Figure \ref{fig:powerdisc} shows the metabolic power demand vs $\gamma$ for the disc shaped tissue, which can be used to identify the optimal $\gamma$. A clear minimum can be found within the range $2<\gamma<4$ in all cases considered. The optimal $\gamma$ is larger than 3 for $\Omega\lesssim 2$ and smaller than 3 for $\Omega \gtrsim 2$. The total metabolic power increases with tree size: For increased $N$, the minimum can be seen to move to slightly higher $\gamma$ for large $\Omega$ and to slightly lower gamma for small $\Omega$, with the optimal $\gamma$ tending slowly towards 3 as the vasculature size is increased. There are only minor differences in the functional form for cases with a single tree or two trees, indicating a lack of dependence of the tree morphology on the number of trees. The anneal parameter used here is $\delta=3\times 10^{-4}$.

\begin{figure*}
    \centering
    \includegraphics[width=75mm]{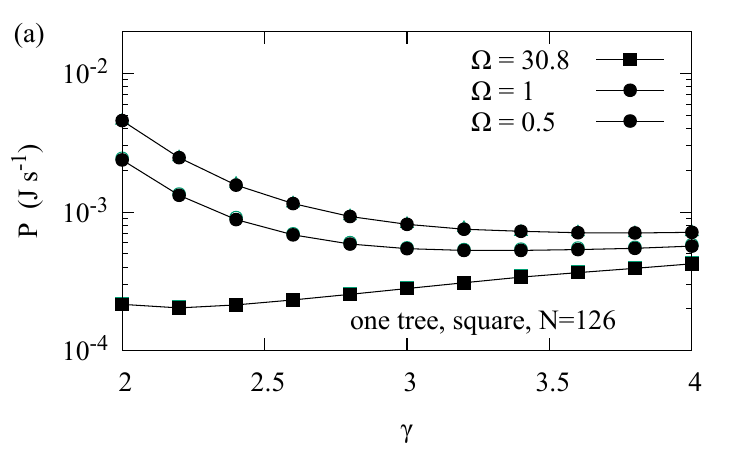}
    \includegraphics[width=75mm]{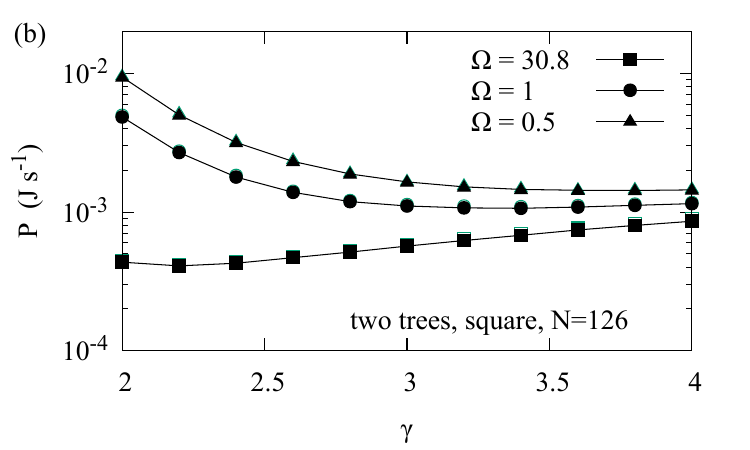}
    \includegraphics[width=75mm]{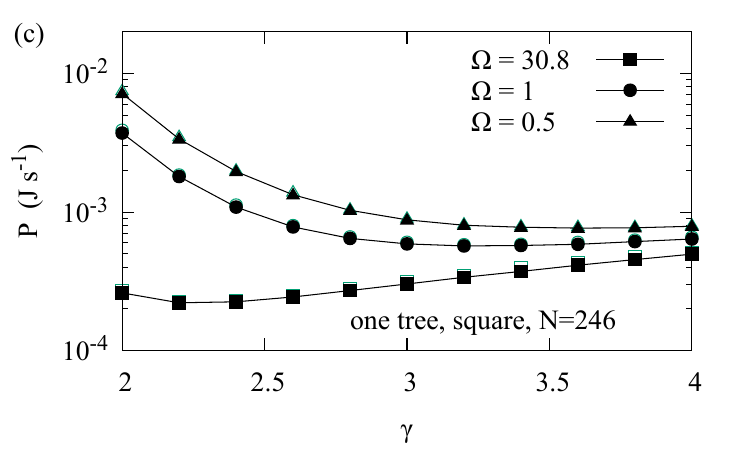}
    \includegraphics[width=75mm]{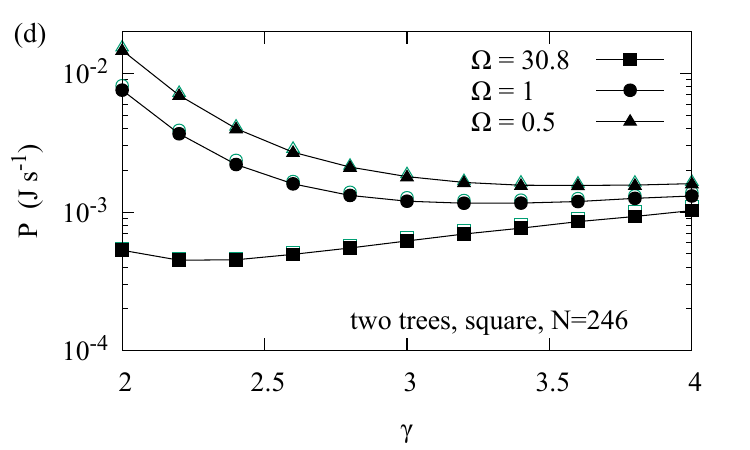}
    \includegraphics[width=75mm]{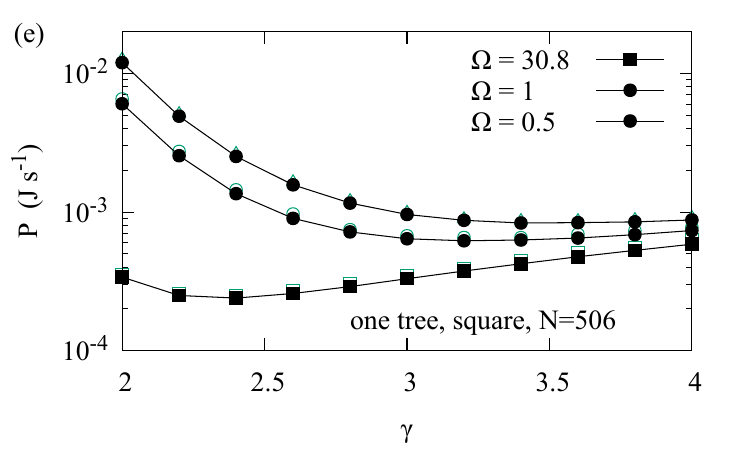}
    \includegraphics[width=75mm]{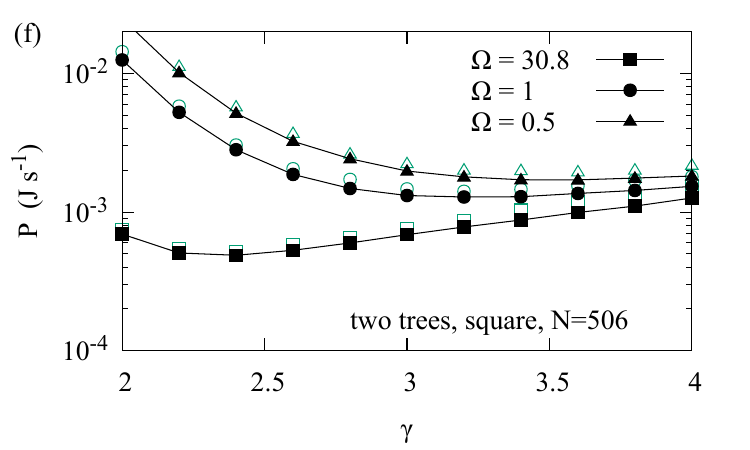}
    \caption{Metabolic power demand vs $\gamma$ for a square of tissue to identify optimal $\gamma$. Similar functional forms for the total metabolic power vs $\gamma$ relationship are found for thin forms of both square and circular tissue, with only small differences in the positions of the minima. Panels show results for (a) a single tree, 126 leaf nodes (b) arterial and venous trees, 126 leaf nodes (c) a single tree, 246 leaf nodes (d) arterial and venous trees, 246 leaf nodes (e) a single tree, 506 leaf nodes (f) arterial and venous trees, 506 leaf nodes. Open turquoise markers show results for shorter anneal time.}
    \label{fig:powersquare}
\end{figure*}

Similar functional forms for the total metabolic power vs $\gamma$ relationship are found for thin forms of both square and circular tissue, with only small differences in the positions of the minima. Figure \ref{fig:powersquare} shows the metabolic power demand, $P$ vs $\gamma$ for a square of tissue, which can be used to identify optimal $\gamma$. Again, minima are essentially independent of the number of trees. For increased $N$, the minimum can be seen to move to slightly higher $\gamma$ for large $\Omega$ and to slightly lower gamma for small $\Omega$. Overall, the optimal values of $\gamma$ do not strongly depend on the shape into which the tissue is grown. The independence of the optimal $\gamma$ from the number of trees is likely related to the independence of bulk morphological properties from tissue shape since the primary interaction between the two trees is the penalty for intersection of segments, which acts to decrease the space available to the other tree (i.e. the shape within which the tree can optimize is modified by the presence of the other tree). The anneal parameter is $\delta=1\times 10^{-4}$. Open turquoise markers show results for shorter anneal time with $\delta=3\times 10^{-4}$. 

\begin{figure*}
    \includegraphics[width=87mm]{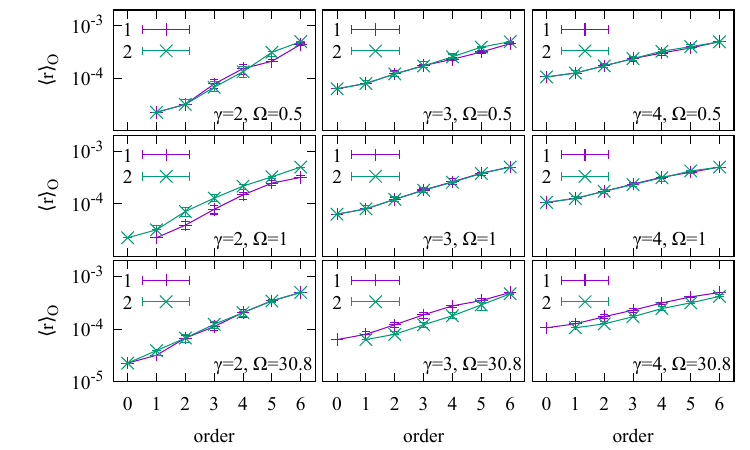}
    \includegraphics[width=87mm]{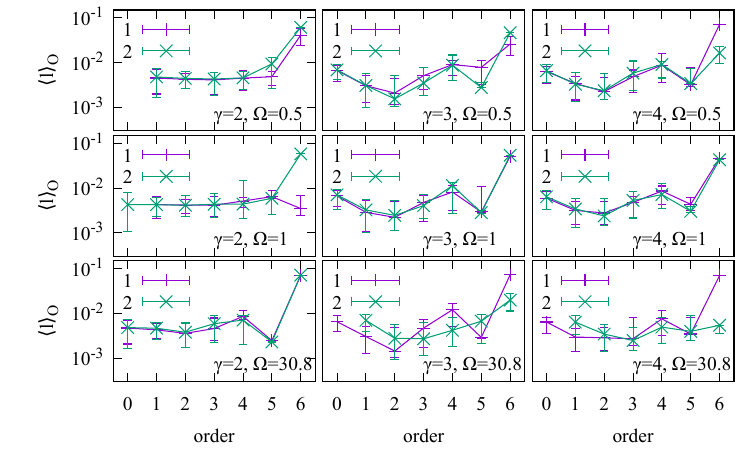}
    \includegraphics[width=87mm]{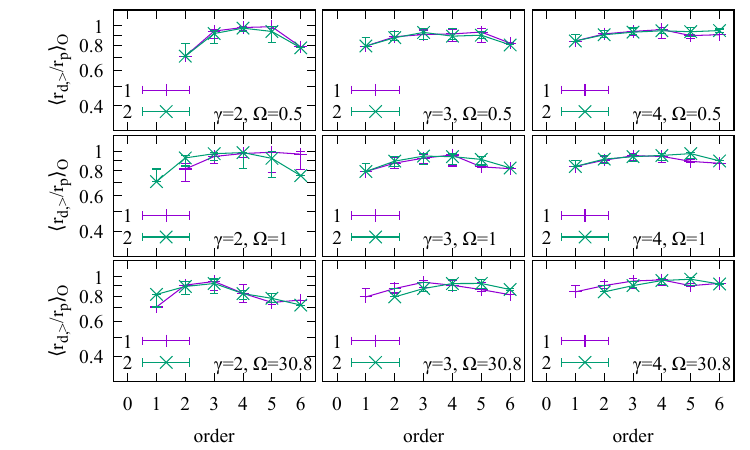}
    \includegraphics[width=87mm]{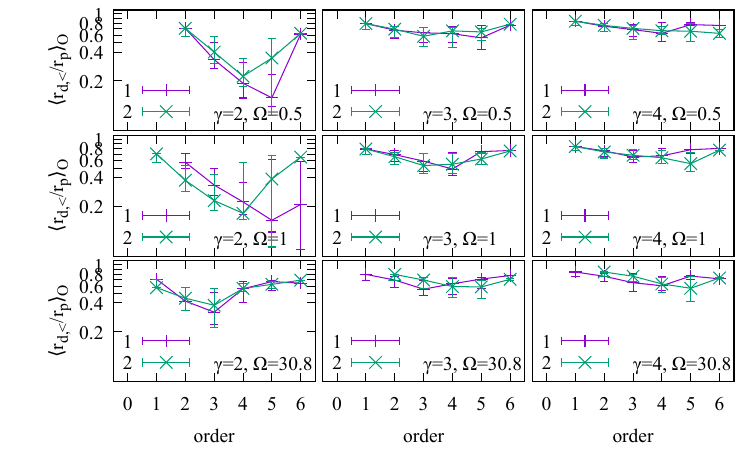}
    \caption{Strahler order analysis of the disc shaped tissue provides a quantitative measure of morphology, showing that morphology is similar in both single and dual tree systems. The average radius $\langle r\rangle_{O}$, length $\langle l\rangle_{O}$, ratio of largest and smallest daughter vessel radii to parent radii $\langle r_{d}/r_{p}\rangle_{O}$ are measured, where the $>$ and $<$ signs indicate largest and smallest daughter vessel respectively. Error bars show 25th and 75th percentile. The largest vessels are labeled with the largest order. }
    \label{fig:strahler}
\end{figure*}

Finally, a Strahler order analysis of the vessels in the disc-shaped tissue is carried out, which shows that the average morphologies of single and dual tree vasculatures is similar (Fig. \ref{fig:strahler}). The Strahler order analysis provides a quantitative overview of morphology to compare one and two tree results. Quantities that are investigated are the average radius $\langle r\rangle$, length $\langle l \rangle$, ratio of largest and smallest daughter vessel radii to parent radii $\langle r_{d}/r_{p} \rangle$, where the $>$ and $<$ signs indicate largest and smallest daughter vessel respectively. The rate at which the average radius decreases with Strahler order changes with $\gamma$ (as expected). For all cases, $\langle l\rangle$ drops suddenly before becoming statistically flat, with the longest vessel lengths associated with the broadest radii. The ratio of the largest daughter radius to the parent radius is typically largest for intermediate Strahler orders. The number of Strahler orders decreases for the largest and smallest $\Omega$ and $\gamma$. The morphology is typically similar for single and dual tree systems, and in the cases where the number of Strahler orders is identical, the morphological properties of the one-tree and two-tree systems are statistically identical. Note that Strahler rather than diameter defined Strahler order is used here in contrast to other publications using SALVO \cite{keelan2019}. While the bulk (average) morphological properties are similar, I emphasise that they are not identical and the detailed structure (particularly the paths of the vessels) can vary significantly between one and two vasculature cases.

\section{Discussion of applications}
\label{sec:discussion}

The method proposed here has applications in quantitative biology and biophysics. To the best of my knowledge there is no alternative scheme to predict the globally optimal structures of multiple vasculatures. The underlying physics of the algorithm enables estimation of the total power expended when pumping blood through the vascular network, which is useful to determine the optimal morphological properties. For example, in this paper, the effect of varying the bifurcation exponent and the metabolic constant have been examined. Moreover, the annealing scheme could be used as a framework for testing and understanding the effects of different contributions from the flow physics and metabolic cost of the vessels. For example, different physical mechanisms could be included to represent the metabolic cost associated with vessel walls, or corrections for pulsatile flow.  Vasculatures grown using this method can be applied to the development of computational models, and SALVO results have already been used in stroke simulations \cite{hague2023a}. 

Moreover, the production of vascularized tissue is a major unsolved problem in biofabrication and tissue engineering, for which the biophysical / quantitative biology based solution proposed here can be applied. Central regions of large engineered tissues cannot obtain the oxygen they require by diffusion, so cells at their centers die without vascuature \cite{schatzlein2022}. The natural vascularization process led by vascular endothelial growth factors and hypoxic gradients is too slow to vascularize large tissues before cell death occurs \cite{mastrullo2020a}. An alternative approach is to decellularize organs to construct a vasculature \cite{wang2022a}, but this is dependent on animal or human donors, and cannot be used to vascularize bespoke tissue volumes.

The approach presented here could be used to design a vasculature for a tissue of arbitrary shape. There are many applications where this capability would be useful: (1) To produce organoids (for applications in drug screening), which can be arbitrarily shaped and require vasculature to emulate real tissue  \cite{zhao2022a}. (2) To create large pieces of cultured meat (e.g. steaks), which requires muscle to be grown that is too thick to receive oxygen and other nutrients by diffusion \cite{schatzlein2022}.  (3) To design an optimal vasculature to supply 3D bioprinted tissue. 3D bioprinting is a key technology for growing cultured tissues that is now approaching the resolution suitable for printing vasculature \cite{chen2021a}. Particularly dense vasculature is required for bioprinted muscle if it is to be transplanted or used for biorobotics \cite{schatzlein2022}.

In summary, there are a variety of applications such as the generation of vasculatures for arbitrary volumes of cultured tissue and cultured meat, for understanding the distribution of blood vessels within organs in quantitative biology, and for use in simulations.

\section{Conclusions}
\label{sec:conclusions}

In this paper, I have introduced an approach for simultaneous determination of the globally optimal arrangement of both arteries and veins required to supply arbitrary tissue volumes. As proof-of-concept, I have demonstrated computational growth of the multiple vasculatures suitable to supply and drain tissues with regularized shapes and compared them with the vessel structures in independent vascular trees. To the best of my knowledge, it is the only approach that is certain to find the global minimum for supply cost for an arbitrary tissue shape.

The core physics of the method is simultaneous optimization of the total power cost for multiple vascular trees: the power required to pump blood through the whole network and the metabolic power cost for maintaining the volume of blood in the vessels. Simulated annealing was applied since it is well suited to combinatorial optimization problems and it is guaranteed to find the global minimum configuration for a sufficiently slow anneal. Simultaneous optimization of both arterial and venous vascular trees is important for the method as it prohibits intersections between tree segments that would short circuit supply to the capillary bed. This differs from previous approximations where venous and arterial trees are optimized independently. The anneal includes updates that make simultaneous changes to multiple vessel segments in order to traverse the configuration space more rapidly. These updates are crucial to mitigate the computational cost associated with checking for intersections between segments in dual vasculatures (and go beyond the two basic updates needed for ergodicity).

In biophysics and quantitative biology, the prediction of metabolically optimal vasculatures is a longstanding problem. The determination of globally optimal vascular trees is important in quantitative biology and biophysics to provide reference points for comparison with evolutionary biology and with other, more approximate, methodologies. Features in the trees are driven by the bifurcation exponent and metabolic constant, which affect whether arteries and veins follow the same or different routes through the tissue, and the level of tortuosity in the vessels. Furthermore, many methods make predictions for only a single vascular tree, yet for many applications it is essential to predict both arterial and venous vasculatures simultaneously to avoid intersections between the vessels that would bypass the capillary network.

A range of potential applications for the method were discussed, including creating vasculatures for 3D bioprinting of cultured tissue and meat, and for use in computational biology. A key application to the algorithm is the determination of vasculatures for 3D printed tissues. Computational design of vasculatures would be an important tool for tissue engineering that will push forward the ability to grow large tissues in combination with 3D bioprinting. For all but the smallest tissues, it is necessary to add a vasculture so that tissue obtains nutrients as soon as it is printed. This demonstration is an important step forward for the computational design of vasculature.

There are a number of possible extensions to the work: These include (1) algorithmic developments to increase the size of the trees that can be predicted, (2) the inclusion of additional physics within the objective function, constraints, and penalties to ensure that vessels follow physiological rules for e.g. bioprinting of tissue.

Algorithmic developments will lead to the optimization of larger vascular trees. When multiple vasculatures are to be optimized, SALVO is limited to relatively small trees with less than $\sim 1000$ segments due to the computational cost of checking for vessel intersections (which grows as the square of the number of segments). Small vascular trees are sufficient to describe small pieces of tissue and small organs (such as the eyes) where the input vessels are narrower and therefore fewer nodes are needed to describe vessels down to the length scales of the capillaries. For larger tissues, alternative optimization methods, such as evolutionary strategies may be needed. Multiscale approaches could also be used to increasing tree size. For example, vascular structure could be determined for small sections of the tissue shape and then tessellated within the tissue.  

Flexibility is a key strength of the approach: by modifying the objective function, new physics and physiology can be added to the simulated annealing framework. Vessel segments that lie outside the tissue can be penalized (which is important for concave tissues and also hollow organs) \cite{keelan2016}. Other constraints can be imposed to exclude large vessels from parenchyma (the functional part of the tissue) \cite{keelan2016,keelan2019}. In some organs, vessels follow very similar paths. For example, in the liver the hepatic artery and portal vein lie close to each other \cite{ABDELMISIH2010643}. To describe such cases, vessels can be constrained to sit within the same region of influence. Furthermore, the method can be extended to include additional vascular and ductal trees. For example, in the liver there are three trees related to vasculature (hepatic artery, hepatic vein and portal vein), a tree of bile ducts (which typically follows the hepatic artery and hepatic vein) and a lymphatic network. Owing to the flexibility of the technique, it is possible to assign different metabolic constants, bifurcation exponents, metabolic power estimates and input radii to each tree.  

Ultimately, I plan to bioprint vasculatures predicted using this technique. This application of the technique is now plausible since 3D bioprinters have achieved the resolution suitable for such printing \cite{chen2021a}.

\section*{Acknowledgements}

I would like to thank Jonathan Keelan and Emma Chung for useful discussions.

\bibliographystyle{unsrt}
\bibliography{references.bib}

\end{document}